# Computational Approach to Investigate Structure-Property Relationship of a series of Carbazole Containing Thermally Activated Delayed Fluorescent Molecules

Md Asif and N. Chawdhury[*]

[1*]Department of Physics, Shahjalal University of Science and Technology, Sylhet-3114, Bangladesh.

*Corresponding author's E-mail(s): nc-phy@sust.edu;

**Abstract**

Donor-acceptor type compounds are an important category of organic materials that show properties suitable for light emission applications. To achieve a full understanding of the mechanism of thermally activated delayed fluorescence (TADF) process, we studied the structure-property relationship for a series of carbazole based TADF emitters, 2CzPN, 4CzPN, 4CzIPN, 4CzBN and 5CzBN. We applied density functional theory to investigate kinetic and electronic properties. We find that the energetic position of triplet excited state of these emitters depends on their molecular structure. Our findings emphasize that to enable reverse intersystem crossing and eventually TADF, strong spin orbit coupling and minimal energy difference between singlet and triplet states $\Delta E_{ST}$ must be obtained simultaneously. We also find that the reverse intersystem crossing rates $k_{RISC}$ values are higher where $\Delta E_{ST}$ values are closer to reorganization energy. Furthermore, a small change in the absorption peak of optical absorption spectra with and without spin orbit coupling (SOC) is observed for each emitter. This result is extremely beneficial for the design of new TADF molecules, and we believe that our work contributes to the progress of future development of high-performance organic molecular light-emitting devices.

**Keywords:** Reverse Intersystem Crossing Rates, Thermally Activated Delayed Fluorescence, Spin Orbit Coupling Matrix Elements, Reorganization Energy

1. Introduction

Organic light-emitting diodes (OLEDs) are a highly researched technology with significant potential for the future of lighting and displays [1-4]. This is due to their extended lifetime, minimal power usage, and notable display developments [5-6]. Following the advancements in OLED technology, a significant breakthrough was achieved in 2012 by Adachi and co-workers, who developed an even more efficient method for light utilization known as Thermally Activated Delayed Fluorescence (TADF), introducing the era of third-generation OLED emitters [7]. These state-of-the-art emitters exhibit exceptional light properties and deliver outstanding device performance. They can achieve near-perfect internal quantum efficiency, meaning nearly all the energy entering the system is converted into usable light [8]. This process involves reverse intersystem crossing (RISC) from the first triplet excited state to the first singlet excited state, facilitated by spin orbit coupling (SOC) between them. The schematic diagram of RISC from triplet to singlet excited states is shown in figure 1.

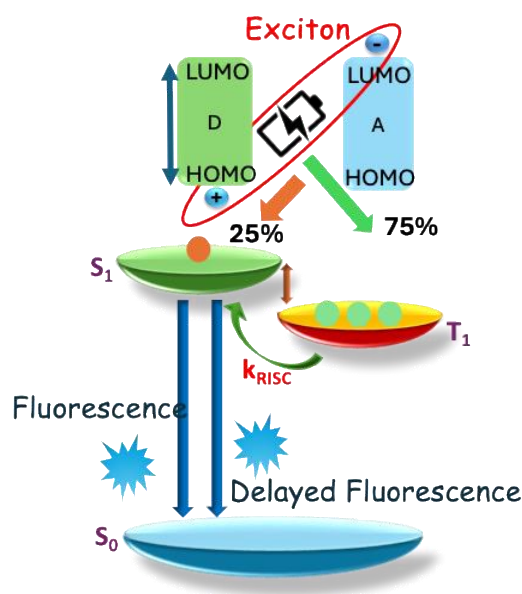

**Figure 1:** Mechanism of Reverse Intersystem Crossing Rate in a Thermally

Donor-acceptor (D-A) organic compounds are an important category of organic materials that may be used in emission applications [9]. Carbazolyl moieties, which act as electron donors, are connected to an electron-accepting dicyanobenzene ring and studied as potential TADF emitters. Steric hindrance causes the carbazolyl units to distort significantly from the plane of the dicyanobenzene ring, creating a distinct separation between the highest occupied molecular orbital (HOMO) and the lowest unoccupied molecular orbital (LUMO). This separation reduces the energy gap ($\Delta E_{ST}$) between the first excited singlet state ($S_1$) and the first excited triplet state ($T_1$), enabling significant spectral overlap between $S_1$ and the triplet manifold ($T_1$–$T_n$) in these carbazole-based TADF emitters [10-13].

The drop in light emission efficiency arises from exciton quenching, likely caused by the slow rate of reverse intersystem crossing ($k_{RISC}$) in the emitter. Furthermore, mechanisms such as singlet-triplet annihilation and triplet-triplet annihilation may also contribute to the quenching. [14]. Scientists are working to overcome the challenges regarding the efficiency of the devices that arises from slow RISC rate.

The light emission mechanism in devices using 4CzIPN as the active material involves a molecule with a benzene ring, where two cyano groups are positioned at the meta-positions as electron-accepting units, and four carbazolyl groups serve as electron-donating units. Their findings suggest that the primary contributor to electroluminescence is the direct recombination of injected electrons and holes. This implies that energy transfer processes play a minimal role in light generation within these devices [15]. Alternatively, another study used 4CzIPN, a compound with a significantly low energy HOMO level, to fabricate a simpler OLED device via a solution-based approach. The incorporation of a hole injection layer with a deep HOMO level, supported by a material that enhances the work function through its self-assembly process, combined with an optimized emission layer, effectively prevents exciton suppression [16].

In a quest to develop solution-processable TADF emitters, researchers strategically modified the existing 4CzIPN. This process involved introducing methyl and t-butyl groups at the (3,6) positions on each of the four carbazole units attached to the central core. This chemical tailoring yielded two novel materials that offered a significant advantage enhanced solubility. Additionally, these modifications contributed to improved film stability. The research team further explored the potential of these materials by fabricating devices using both solution-based and vacuum processing techniques [17].

A separate study explored the impact of steric hindrance from t-butyl groups on the shielding effect in TADF materials. The researchers used four carbazolyl groups evenly arranged on the benzene ring adjacent to a nitrile group (4CzBN) and five carbazolyl groups similarly positioned evenly on the benzene ring (5CzBN). They then introduced t-butyl groups, creating sterically hindered variants: 4TCzBN and 5TCzBN. This modification resulted in a slight reduction in the energy gap between the singlet and triplet excited states ($\Delta E_{ST}$), while also increasing the oscillator strength [18].

To improve the performance of blue-emitting TADF materials in both vacuum and solution-processed devices, a novel material called 5CzCN has been studied. This innovative material features benzonitrile as the electron-accepting unit and five carbazole units as the electron-donating components. Their approach led to remarkable results, achieving a maximum external quantum efficiency (EQE) of 19.7% in TADF devices fabricated using vacuum deposition and 18.7% using solution processing techniques [19]. Additionally, researchers modified the green-emitting 4CzIPN by replacing one nitrile group with a carbazole unit to create 5CzBN, which emits in the blue wavelength [19, 20].

In this study, we present theoretical calculations on a series of carbazole-based donor-acceptor molecules to explore the relationship between their kinetic properties and molecular structures. We calculate the RISC rates for a range of TADF emitters with carbazole units as donor, 4,5-Bis(carbazol-9-yl)-1,2-dicyanobenzene (2CzPN), 3,4,5,6 tetrakis(carbazol-9-yl)-1,2-dicyanobenzene (4CzPN), 1,2,3,5 tetrakis(carbazol-9-yl)-4,6-dicyanobenzene (4CzIPN), 2,3,5,6 tetrakis(carbazol-9-yl) benzonitrile (4CzBN) and 2,3,4,5,6-penta (carbazol-9 yl) benzonitrile (5CzBN). The RISC rate constant, $k_{RISC}$, can be calculated by using Marcus- Hush equation [21-23]:

$$k_{RISC} = \frac{2\pi}{\hbar}|V_{SOC}|^2 \times \rho_{FCWD} \qquad (1)$$

Here, $k_{RISC}$ is directly proportional to $|V_{SOC}|^2$ and it is the coupling term of the Hamiltonian $\widehat{H_{SOC}}$. In transition from $T_1$ to $S_1$, where $T_1$ has three spin sublevels with eigen values $m_S = -1, 0, +1$. Therefore, to get the value of $|V_{SOC}|^2$ one takes the average of three spin-orbit coupling matrix elements (SOCME) between $T_1$ and $S_1$:

$$|V_{SOC}|^2 = \frac{1}{3}\left|\langle S_1|\widehat{H_{SOC}}|T_1\rangle\right|^2$$

Frank-Condon-weighted density of states $\rho_{FCWD}$ is another important factor to calculate RISC rates, that is explained simply by the classical Marcus theory,

$$\rho_{FCWD} = \frac{1}{\sqrt{4\pi\lambda k_B T}} exp\left(-\frac{(\Delta E_{ST}+\lambda)^2}{4\lambda k_B T}\right) \qquad (2)$$

Where λ is the reorganization energy for the respective transition, and $k_B$ is Boltzmann constant.

Density functional theory has been utilized to obtain the values of $\Delta E_{ST}$ and SOCME of the investigated molecules. To investigate the spin-flip processes in the TADF mechanism we applied the above powerful equations.

## 2. Computational Methods

In this work, we employ computational methods to explore the energetics of singlets ($S_1$, $S_2$ and $S_3$) and triplets ($T_1$ and $T_2$), HOMO and LUMO levels, SOCME, dihedral angle between the electron donor and acceptor, optical absorption, and RISC rates of the materials. Gaussian16 and ORCA software packages were used for the geometry optimization and calculations of the energies [24]. Ground state geometry optimizations were performed with the B3LYP functional and DEF2-SVP basis set under the density functional theory (DFT) framework to determine the most stable electronic structure of the materials. Figure 2 shows the chemical and optimized structures of five carbazole based TADF emitters.

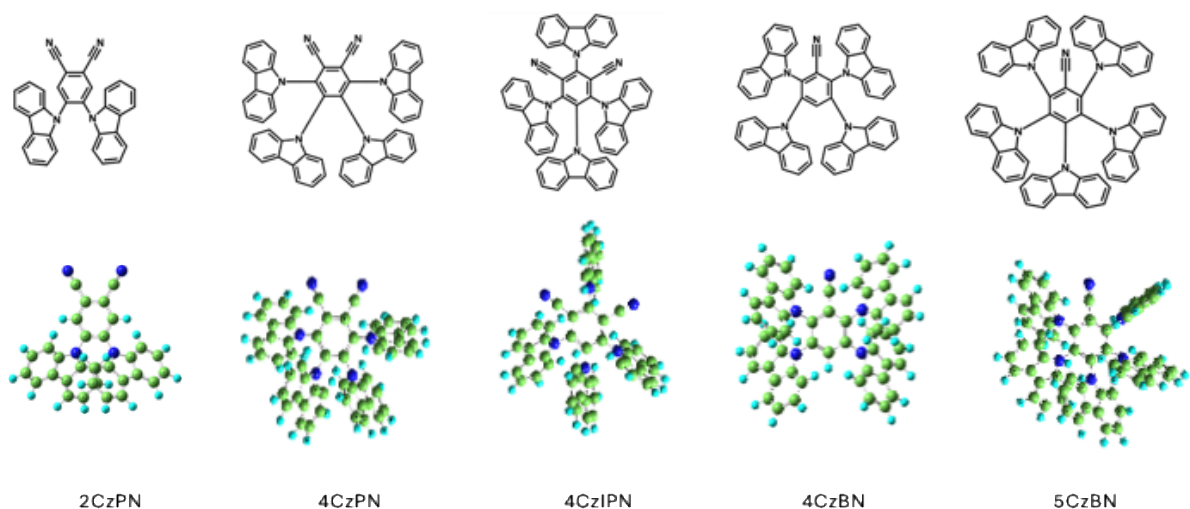

**Figure 2:** Chemical and optimized structures of five carbazole based TADF emitters.

HOMOs, LUMOs and optical absorption spectra were obtained using DFT and time dependent DFT respectively. Excited singlet and triplet state energies and SOCME were calculated in the environment of methyl chloride solvent and using time-dependent DFT with ORCA software. In this case, the B3LYP functional was employed, with a different basis set, ZORA-DEF2-TZVP. For further accuracy, two different values of reorganization energy, $\lambda = 0.1$ eV and $\lambda = 0.2$ eV, were used to calculate the RISC rates.

## 3. Results and Discussion

### 3.1 HOMOs, LUMOs and Optical Absorption Spectra

Figure 3 illustrates the HOMOs and LUMOs of all the molecules. It is observed that the HOMOs are predominantly delocalized over the carbazolyl moieties, while the LUMOs are mainly localized on the benzonitrile or dicyanobenzene moieties. This suggests that the carbazolyl units function as electron donors, whereas the benzonitrile and dicyanobenzene units serve as electron acceptors. All five molecules exhibit twisted structures, with dihedral angles ranging from 118° to 127° between the planes of the carbazolyl and benzonitrile or dicyanobenzene groups. This twisting is a result of significant steric hindrance between the carbazolyl and benzonitrile or dicyanobenzene units. In the case of 5CzBN, where the number of carbazolyl groups is increased, the HOMO is not fully delocalized over all five carbazolyl units.

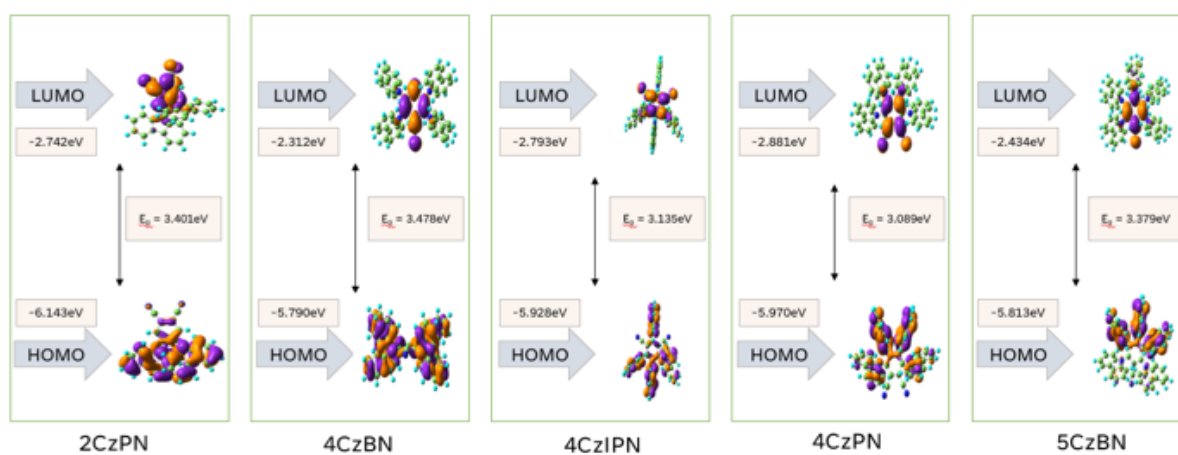

**Figure 3**: Calculated HOMO, LUMO, their energies and HOMO-LUMO energy gaps of Carbazole Based TADF.

Consequently, we find that the HOMOs and LUMOs are well separated with the HOMO-LUMO energy gaps of 3.40 eV, 3.48 eV, 3.14 eV, 3.09 eV and 3.38 eV for 2CzPN, 4CzBN, 4CzIPN, 4CzPN and 5CzBN, respectively. Because of well separated HOMOs and LUMOs, small $\Delta E_{ST}$ values can be anticipated which are mandatory for high RISC rates [25].

Normalized optical absorption spectrum of five carbazole based molecules are shown in figure 4. All the molecules absorb in the blue region of the solar radiation with the absorption peak of 2CzPN, 4CzBN, 4CzIPN, 4CzPN and 5CzBN are at 3 eV, 2.83 eV, 2.64 eV, 2.5 eV, 2.81 eV respectively. Previous measurement on optical absorption of 4CzIPN reported two a peak and a shoulder at 3.35 eV and 2.76 eV, respectively. However, our theoretical optical spectrum shows only one peak around at 3.35 eV. The photoluminescence of this molecule is between 2.48 eV and 2.25 eV depending on the solvent used [26]. The measured optical absorption peak of 4CzBN and 5CzBN are around 2.95 eV which are 0.12 eV and 0.14 eV lower than our calculated values of 2.83 eV and 2.81 eV, respectively. These absorption bands are primarily attributed to electron transfer from the carbazolyl unit to the benzonitrile unit. The two molecules exhibit blue emission in toluene, with emission peaks at 2.81 eV for 4CzBN and 2.67 eV for 5CzBN, respectively [18]. The absorption spectra show a slight red shift upon introducing dicyanobenzene substituents, as observed for 4CzPN and 4CzIPN. However, when the number of carbazolyl groups is reduced, as in 2CzPN, the absorption spectrum is blue-shifted (Figure 5). We also accounted for spin-orbit coupling in the absorption spectra calculations and observed no significant change, due to the very low SOCME values less than 1.0 cm$^{-1}$ for our TADF materials.

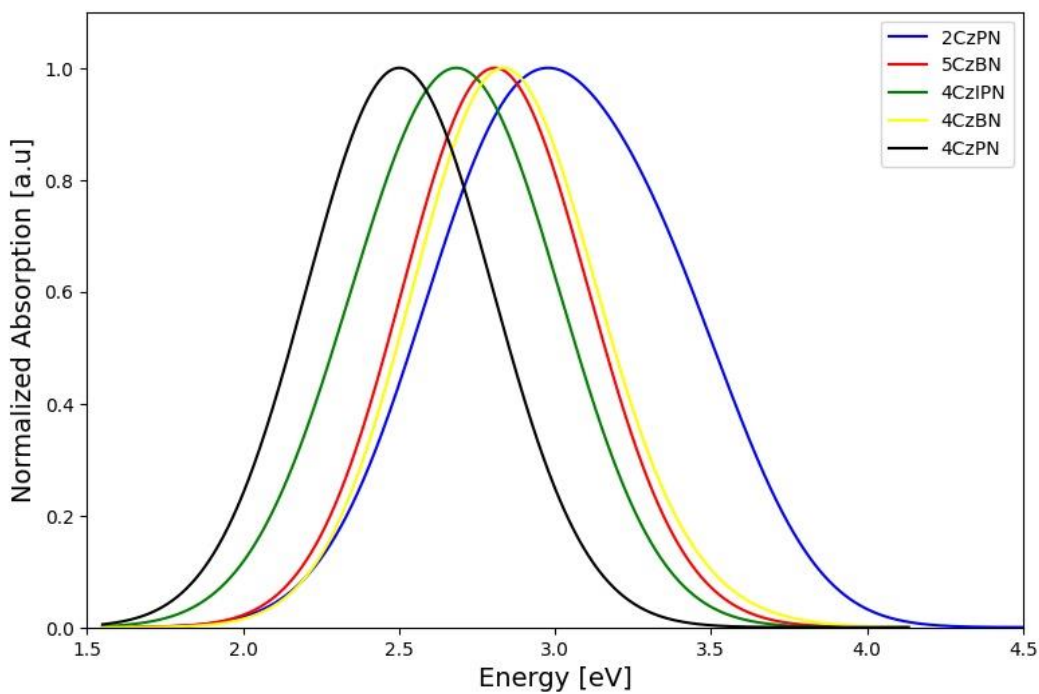

**Figure 4**: Optical Absorption spectra of Carbazole based TADF emitters.

### 3.2 Energy Levels and Reverse Intersystem Crossing Rates

The energy between lowest singlet and lowest triplet excited state is an important parameter for efficient reverse intersystem crossing. Therefore, we have computed the energies of lower lying singlet and triplet states. For comparison, the energy level diagram of the emitters is shown in Figure 5. Our calculation shows that the energetic positions of singlets and triplets depends on the number of carbazole units in the molecules. For example, there are two triplet excited states, $T_1$ and $T_2$ below the first singlet excited state $S_1$ in the molecules 4CzBN, 4CzIPN and 4CzPN where there are four carbazole units. However, there are one and three triplet excited state are below S1 for 2CzPN and 5CzBN, respectively. These results distinctly suggest a structure-property relationship between the carbazole containing emitters and their electronic structure. The summary of the calculated kinetic properties i.e. RISC rates and energetics of the emitters at 300K including the spin-orbit coupling matrix elements are given in table 1.

**Table 1:** Energetics of lower lying singlet (S1 and S2) and triplet (T1, T2, and T3) excited states and energy gap between lowest excited singlet and lowest excited triplet states ($\Delta E_{ST}$), Spin-Orbit Coupling Matrix Element (SOCME), RISC rates ($k_{RISC}$) both calculated for $\lambda = 0.1$ eV and $\lambda = 0.2$ eV and dihedral angle, $\phi$.

| Emitter | S1 (eV) | S2 (eV) | T1 (eV) | T2 (eV) | T3 (eV) | $\Delta E_{ST}$ (eV) | SOCME ($cm^{-1}$) | $k_{RISC}$ x $10^5$ ($s^{-1}$) $\lambda$ = 0.1eV | $k_{RISC}$ x $10^5$ ($s^{-1}$) $\lambda$ = 0.2 eV | $k_{RISC}$ x $10^5$ ($s^{-1}$) (Experimental) | $\phi$ (degree) |
|---|---|---|---|---|---|---|---|---|---|---|---|
| 2CzPN | 2.66 | 2.90 | 2.40 | 2.68 | 2.83 | 0.26 | 0.51 | 0.003 | 0.02 | 0.05{30] | 127 |
| 4CzPN | 2.43 | 2.44 | 2.24 | 2.32 | 2.43 | 0.18 | 0.40 | 0.188 | 0.26 | | 119 |
| 4CzIPN | 2.47 | 2.66 | 2.35 | 2.43 | 2.60 | 0.12 | 0.22 | 1.38 | 0.72 | 20.3[29] | 119 |
| 4CzBN | 2.81 | 3.03 | 2.67 | 2.75 | 2.99 | 0.15 | 0.27 | 0.59 | 0.44 | 1.8[29] | 122 |
| 5CzBN | 2.76 | 2.76 | 2.54 | 2.64 | 2.73 | 0.22 | 0.99 | 0.15 | 0.41 | 2.4[29] | 118 |

We observe lowest $\Delta E_{ST}$ is shown for 4CzIPN, on the other hand the highest SOCME value is observed for 5CzBN. This effect is attributed to increased number of heavy nitrogen atoms.

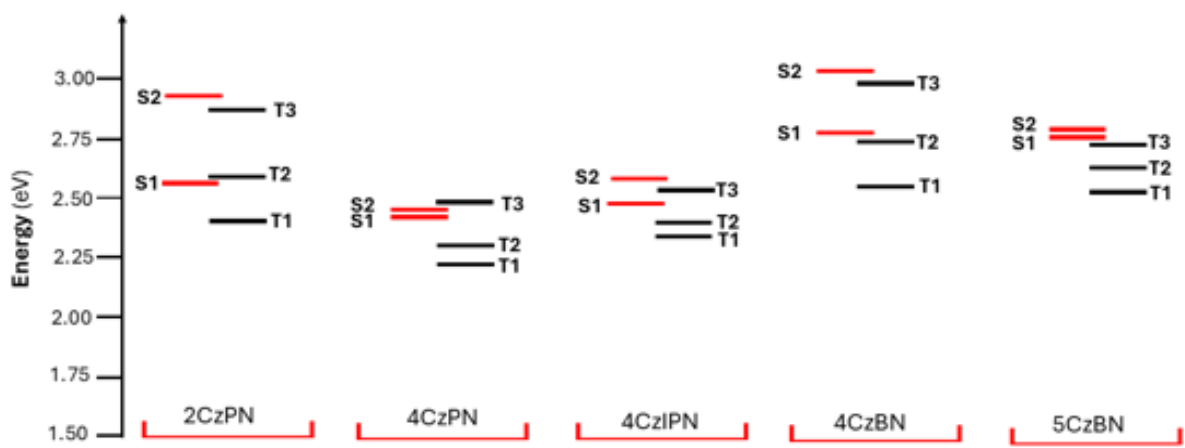

**Figure 5**: Energy-level diagram of the excited states of investigated emitters.

Previous reports suggest that the reorganization energy, $\lambda$, for purely organic molecules lie between 0.1 eV to 0.2 eV [27, 28]. Therefore, we have calculated $k_{RISC}$ with $\lambda$ = 0.1eV and $\lambda$ = 0.2eV at 300 K for all five molecules. Figure 6 shows variation of $k_{RISC}$ with $\Delta E_{ST}$ with both the values of $\lambda$ for comparison. We find that 4CzIPN has the maximum value of $k_{RISC}$ with $\lambda$ = 0.1eV which is comparable to the value of $\Delta E_{ST}$ of 0.12eV. For each molecule $k_{RISC}$ value is higher where $\Delta E_{ST}$ value is closer to $\lambda$. For instances $k_{RISC}$ of 4CzPN, 5CzBN and 2CzPN are higher for $\lambda$ = 0.2 eV where $\Delta E_{ST}$ (= 0.18 eV, 0.22 eV and 0.26 eV, respectively) of these molecules are comparable to the value of reorganization energy.

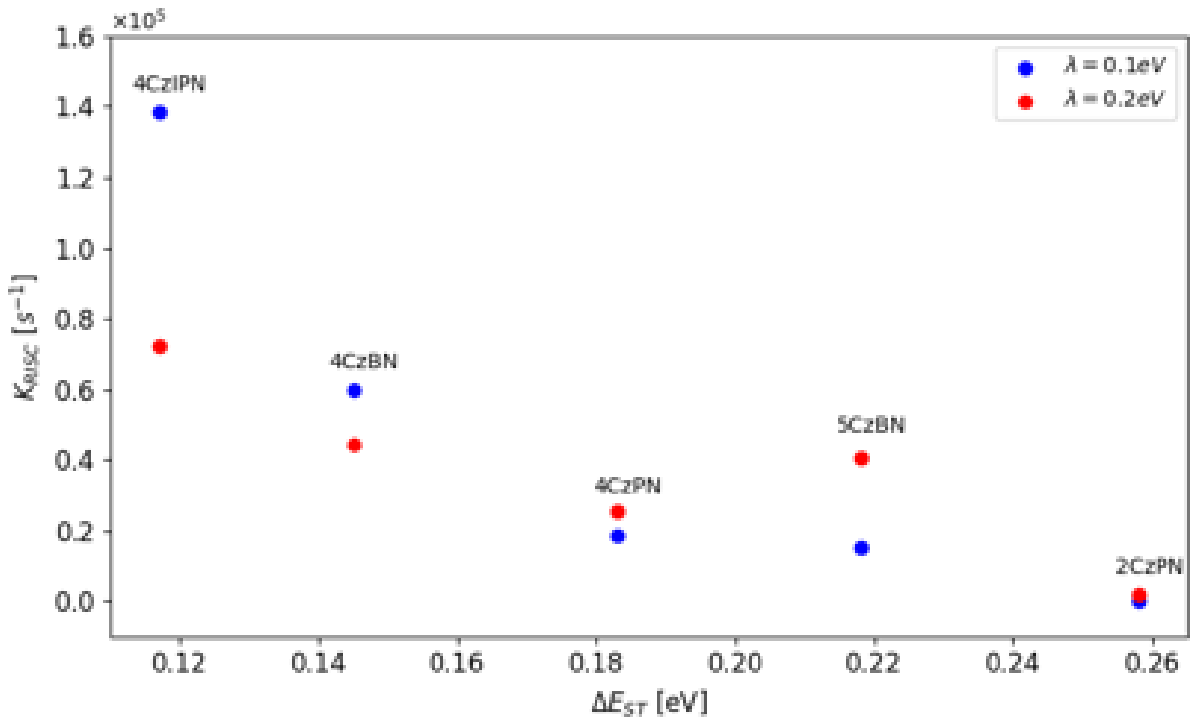

**Figure 6**: Variation of calculated $k_{RISC}$ with $\Delta E_{ST}$ for $\lambda = 0.1$eV and $\lambda = 0.2$eV at 300K.

The available experimental values of $k_{RISC}$ are shown in table 1 for comparison with our calculated $k_{RISC}$ values of the molecules [29,30]. Hosokai et. al. proposed a chemical structure that suppress the structural relaxation by introducing bulky moieties around linearly positioned donor units in a donor-acceptor-donor (D-A-D) structure. In their study they assigned 4CzIPN, 4CzBN and 5CzBN as D-A-D, and 2CzPN and 4CzPN as D-A structured molecules. In our study, we find that the $k_{RISC}$ of D-A-D structured molecules are greater than D-A structured molecules. This finding is consistent with Hosokai et. al. proposal. [29] estimated the RISC rates of twenty D-A and D-A-D structured molecules using experimental data for rate constants of prompt fluorescence, delayed fluorescence, and photoluminescence quantum yields for both types of fluorescence from the literature. They investigated TADF materials and classified them based on their $k_{RISC}$ values. For 5CzBN and 4CzIPN, the $k_{RISC}$ values were an order of magnitude lower than their experimental counterparts. However, molecules like 4CzBN, 2CzPN, and 4CzPN exhibited similar $k_{RISC}$ values [31]. The Monkman group explored a TADF emitter with a D-A-D structure, where the D and A subunits are almost orthogonally oriented. They proposed that the orientation of the D-A units plays a crucial role in the reverse intersystem crossing processes, contributing to high TADF efficiency [32]. Furthermore, they examined how different environments, such as solution and solid state, affect the energy splitting between singlet and triplet states in a D-A type TADF emitter [33]. Singlet and triplet excited state calculations were performed in the presence of methyl chloride solvent.

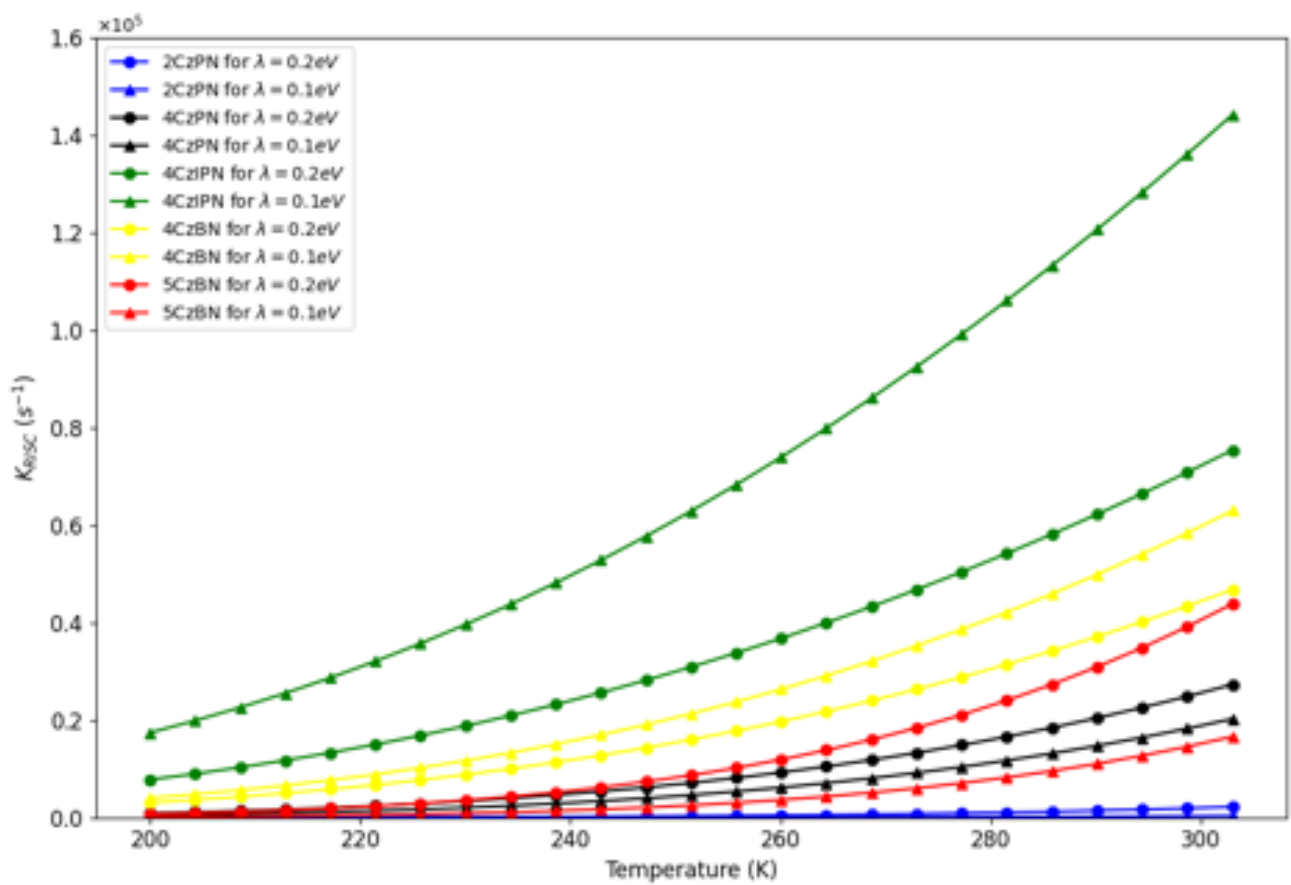

**Figure 7:** Temperature Dependence of $k_{RISC}$.

According to Marcus-Hush equation the RISC rate can be improved by increasing the thermal energy by raising temperature. Therefore, it is necessary to study the temperature dependence of $k_{RISC}$ of the molecules. (Figure 7) We note that, 4CzIPN shows strong dependency on temperature. However, 2CzPN has an insignificant change with increasing thermal energy. We attribute this behavior as non-TADF property.

## 4. Conclusion

In this computational work, we systematically examined the interplay between molecular structure, kinetic property, and electronic structure of a series of carbazole containing donor-acceptor type molecules. We find that the energetic position of triplet excited state of these emitters depends on their molecular structure. The wide separation of HOMOs and LUMOs indicate that these molecules are well stable and their singlet-triplet separation is relatively lower. For understanding the optical behavior, we simulate optical spectrum with and without considering spin orbit coupling. From the optical spectra we observe that there is a slight change in the absorption peak when we consider spin orbit coupling. For kinetic properties, we calculated reverse intersystem crossing rates. Our calculated reverse intersystem crossing rates are consistent with the available experimental data.

In conclusion we find the following conditions to meet simultaneously for achieving faster reverse intersystem crossing rates: strong spin-orbit coupling, narrow singlet-triplet energy gap and the reorganization energy should be close to the gap between first singlet and first triplet excited states. This result is extremely beneficial for the design of new TADF molecules, and we believe that our work contributes to the progress of future development of high-performance organic molecular light-emitting devices.

**References**

[1] H.-H. Cho, et al., *Nat. Photonics* **2024**, *1*, 1–8.

[2] X.-K. Chen, D. Kim, J.-L. Brédas, *Acc. Chem. Res.* **2018**, *51*, 2215–2224.

[3] M. Y. Wong, E. Zysman-Colman, *Adv. Mater.* **2017**, *29*, 1605444.

[4] K. Shizu, H. Kaji, *Nat. Commun.* **2024**, *15*, 4723.

[5] L.-S. Cui, A. J. Gillett, S.-F. Zhang, H. Ye, Y. Liu, X.-K. Chen, Z.-S. Lin, et al., *Nat. Photonics* **2020**, *14*, 636–642.

[6] S. S. Sudheendran, S. Sujith, et al., *Adv. Sci.* **2021**, *8*, 2002254.

[7] Q. Zhang, B. Li, S. Huang, H. Nomura, H. Tanaka, C. Adachi, *Nat. Photonics* **2014**, *8*, 326–332.

[8] T. Miwa, et al., *Sci. Rep.* **2017**, *7*, 284.

[9] Y. Olivier, M. Moral, L. Muccioli, J.-C. Sancho-García, *J. Mater. Chem. C* **2017**, *5*, 5718–5729.

[10] K. Shizu, J. Lee, H. Tanaka, H. Nomura, T. Yasuda, H. Kaji, C. Adachi, *Pure Appl. Chem.* **2015**, *87*, 627–638.

[11] F. Dumur, *Org. Electron.* **2015**, *25*, 345–361.

[12] N. Blouin, M. Leclerc, *Acc. Chem. Res.* **2008**, *41*, 1110–1119

[13] T. Zhang, et al., *Angew. Chem. Int. Ed.* **2023**, *62*, e202301896.

[14] J. W. Sun, K.-H. Kim, C.-K. Moon, J.-H. Lee, J.-J. Kim, *ACS Appl. Mater. Interfaces* **2016**, *8*, 9806–9810.

[15] P. Wang, S. Zhao, Z. Xu, B. Qiao, Z. Long, Q. Huang, *Molecules* **2016**, *21*, 1365.

[16] Y.-H. Kim, C. Wolf, H. Cho, S.-H. Jeong, T.-W. Lee, *Adv. Mater.* **2015**, *28*, 734–741.

[17] Y. Im, M. Kim, Y. J. Cho, J.-A. Seo, K. S. Yook, J. Y. Lee, *Chem. Mater.* **2017**, *29*, 1946–1963.

[18] D. Zhang, M. Cai, Y. Zhang, D. Zhang, L. Duan, *Mater. Horiz.* **2016**, *3*, 145–151.

[19] Y. J. Cho, S. K. Jeon, J. Y. Lee, *Adv. Opt. Mater.* **2016**, *4*, 688–693.

[20] S. Tanimoto, T. Suzuki, H. Nakanotani, C. Adachi, *Chem. Lett.* **2016**, *45*, 770–772.

[21] R. A. Marcus, *Angew. Chem. Int. Ed.* **1993**, *32*, 1111–1121.

[22] M. Mońka, I. E. Serdiuk, K. Kozakiewicz, E. Hoffman, J. Szumilas, A. Kubicki, ... P. Bojarski, *J. Mater. Chem. C* **2022**, *10*, 7925–7934.

[23] I. E. Serdiuk, *J. Phys. Chem. B* **2021**, *125*, 2696–2706

[24] F. Neese, *Wiley Interdiscip. Rev.: Comput. Mol. Sci.* **2012**, *2*, 73–78.

[25] S. Hirata, Y. Sakai, K. Masui, H. Tanaka, S. Y. Lee, H. Nomura, ... C. Adachi, *Nat. Mater.* **2015**, *14*, 330–336.

[26] R. Ishimatsu, S. Matsunami, K. Shizu, C. Adachi, K. Nakano, T. Imato, *J. Phys. Chem. A* **2013**, *117*, 5607–5612.

[27] K. Schmidt, S. Brovelli, V. Coropceanu, D. Beljonne, J. Cornil, C. Bazzini, T. Caronna, R. Tubino, F. Meinardi, Z. Shuai, et al., *J. Phys. Chem. A* **2007**, *111*, 10490–10499.

[28] D. Beljonne, Z. Shuai, G. Pourtois, J. Bredas, *J. Phys. Chem. A* **2001**, *105*, 3899–3907.

[29] T. Hosokai, H. Matsuzaki, H. Nakanotani, K. Tokumaru, T. Tsutsui, A. Furube, K. Nasu, H. Nomura, M. Yahiro, C. Adachi, *Sci. Adv.* **2017**, *3*, 1603282.

[30] T. Furukawa, H. Nakanotani, M. Inoue, C. Adachi, *Sci. Rep.* **2015**, *5*, 8429.

[31] F. B. Dias, J. Santos, D. R. Graves, P. Data, R. S. Nobuyasu, M. A. Fox, ... A. P. Monkman, *Adv. Sci.* **2016**, *3*, 1600080.

[32] N. Aizawa, Y. Harabuchi, S. Maeda, et al., *Nat. Commun.* **2020**, *11*, 3909

[33] P. L. Santos, J. S. Ward, P. Data, A. S. Batsanov, M. R. Bryce, F. B. Dias, A. P. Monkman, *J. Mater. Chem. C* **2016**, *4*, 3815.